# VERIFICATION OF CONVERGENT-DIVERGENT NOZZLE DESIGNS IN PROPULSION AEROSPACE APPLICATIONS

**Noah L. Estrada[1], Marc A. Cantu[1], Arturo Rodriguez[1*], Andrew R. Ybarra[1], Jeffrey H. Farris[1], Francisco O. Aguirre Ortega[1], Vineeth Vijaya Kumar[1], Vinod Kumar[1]**

[1]Texas A&M University - Kingsville, Kingsville, TX 78363, USA

## ABSTRACT

The performance of convergent–divergent nozzles is critical in aerospace propulsion systems, where the efficient expansion of high-temperature, high-pressure gases directly impacts thrust generation. In this study, we investigate a series of nozzle geometries using numerical simulations in ANSYS Fluent, guided by classical compressible flow theory, initially developed by Ludwig Prandtl. The governing equations of conservation of mass, momentum, and energy are solved under steady-state conditions, with emphasis on shock formation, boundary-layer effects, and Mach number distributions across the nozzle throat and divergent section. Parametric analyses are conducted to evaluate the influence of nozzle contour, area ratio, and throat geometry on flow acceleration and thrust coefficient. The results demonstrate close agreement with theoretical predictions of isentropic compressible flow while also highlighting deviations due to viscous and three-dimensional effects. These findings provide design insights for optimizing nozzle performance across propulsion applications, from launch vehicles to high-speed air-breathing systems. We obtained absolute error differences of 2.05%, 6.03%, and 9.9% in the throat temperature measurements for the RL10B2, SSME-40k, and A-1 nozzles, respectively.

## NOMENCLATURE

**Variables**

| | | |
|---|---|---|
| $\rho$ | density | kg/m$^3$ |
| $P$ | pressure | Pa |
| $R$ | gas constant | J/kg·K |
| $T$ | temperature | K |
| $M$ | mach number | |
| $V$ | velocity | m/s |
| $c$ | speed of sound | m/s |
| $k$ | specific heat ratio | |

**Subscripts and Superscripts**

| | |
|---|---|
| 0 | inflow condition |
| * | throat |
| f | function |

## INTRODUCTION

The convergent-divergent (C-D) nozzle remains a standard configuration for compressible flow studies, providing insight into the interplay among fluid acceleration, shock formation, and boundary-layer development [1, 2]. Despite its simplicity, the C-D nozzle encompasses all the essential elements of modern computational fluid dynamics verification, including conservation equations, discretization, and fidelity, as well as the need to compare analytical solutions with numerical simulations [3, 4]. In the context of hypersonic flow and propulsion system design, CFD predictions require a fundamental verification exercise in defined geometries to extend turbulent, viscous, and chemically reacting flow models [5]. This work focuses on the verification of CFD solutions for a C-D nozzle against analytical solutions derived from isentropic flow relations and shock-expansion theory [6]. By establishing correspondence between the numerical and analytical solutions for accuracy, Mach number, and distributions across the nozzle, the study demonstrates the solver's internal consistency and convergence behavior.

*Corresponding Author: Arturo.Rodriguez@tamuk.edu

The motivation for such verification extends beyond academic interest. In the design of wind tunnels, rocket engines, and hypersonic flow test facilities, numerical discrepancies at the flow exit can lead to errors in stagnation enthalpy and flow uniformity [7, 8, 9, 10, 11]. Through verification—combining analytical theory, steady-state CFD, and fine-meshing—the presented work contributes to establishing a robust validation of compressible internal flows. Ultimately, the convergent-divergent nozzle serves pedagogically as a technical comparison, reinforcing verification and numerical reproduction.

C-D nozzles, or Laval nozzles, are fundamental to aerospace propulsion, converting the thermal and potential energy of combustion gases at high pressure into kinetic energy for propulsion [11]. The converging section accelerates the flow to sonic speed; this is followed by the diverging section, which expands the flow to supersonic speed. These nozzles optimize high-speed performance in rocket engines, supersonic jet applications, and advanced wind tunnels. Design verification of C-D nozzles is critical in modern aerospace engineering to ensure reliability, operational safety, and efficiency under extreme temperatures, as well as the accuracy and flow conditions typical of launch vehicles, advanced aircraft, and propulsion system testing [12].

The efficiency performance of C-D nozzles directly contributes to engine thrust and efficiency by achieving supersonic exit velocities [13]. However, the challenges of shock waves, flow separation, and material stresses require rigorous verification, combined with analytical modeling, numerical CFD simulations, and experimental testing, to prevent propulsion loss, fuel consumption, or structural failure [14]. The paper details the integration of C-D nozzle design verification, focusing on performance optimization and robust validation for aerospace systems development. By supporting model comparisons, sensitivity analysis, and advanced CFD tools, the study advances nozzle design methodologies.

## LITERATURE REVIEW

The C-D nozzle literature focuses on propulsion optimization design, flow characterization, material innovation, and verification techniques. Fundamental research on these topics began at NASA and other similar agencies, focusing on the design and relationship between nozzle geometries and thrust coefficients. The control figure, which reaches a thrust coefficient of 0.985, can be further improved by minimizing coolant leakage and loss, as highlighted [13]. Numerical simulations, particularly computational fluid dynamics simulations, support these studies to optimize nozzle designs; generative algorithms and numerical solvers also refine geometry parameters effectively for applications ranging from wind tunnels to rocket vehicles [14].

Multiple investigations examine the impact of convergence and divergence angles, demonstrating that large divergence angles increase the risk of separation and reduce efficiency; careful angle selection improves thrust, exit velocity, and temperature management [15]. Structural studies comparing Titanium Carbide and Inconel 718 demonstrate the materials' performance at varying temperatures and their accuracy during propulsion, guiding material selection for high-temperature, high-pressure applications [16]. Verification methodologies include comparisons between numerical and analytical models, demonstrating the importance of verification for accurately predicting the physics of high-temperature gas dynamics [14]. Experimental and numerical studies analyze nozzle area variability, two-phase flows under air injection, and their optimization using Taguchi and U.S. Department of Energy methods, providing strategies for improving throat diameters and angles to maximize mass flow and thrust [17].

Recent literature has improved verification methods by combining experimental data, computational fluid dynamics predictions, and theoretical models to bridge the gap between idealized numerical simulations and real-world performance [14]. Prolonged implementation has left persistent gaps in high-fidelity modeling and real-time validation, reinforcing the need for coordinated verification strategies for aerospace propulsion systems.

## THEORETICAL REVIEW

The operation of C-D nozzles is governed by compressible flow theory, primarily through the relationships of one-dimensional isentropic flow and expansion [12]. The Laval-shaped nozzle dictates the acceleration of subsonic flow in the convergence duct, reaching sonic velocity (Mach 1) at the throat, and expanding to supersonic velocity in the divergence section, governed by the area-Mach number relationship in the isentropic flow equations [12]:

**Table 1** Isentropic Compressible Flow Equations

|  |  |
|---|---|
| $\rho_0 = \frac{P_0}{RT_0}$ | (1) |
| $\frac{P^*}{P_0} = f(M)$ | (2) |
| $\frac{T^*}{T_0} = f(M)$ | (3) |
| $\frac{\rho^*}{\rho_0} = f(M)$ | (4) |
| $V^* = c^* = \sqrt{kRT^*}$ | (5) |

The assumptions of isentropic flow are adjusted for practical aerospace operations, where shock waves, boundary layer separation, and viscous and thermal losses occur. Divergence factor corrections are considered for force loss due to real-world effects [13]. Increasing the use of CFD in modeling heat transfer, viscous phenomena, and design behavior will help verify theoretical predictions [14]. The robust framework supports the systematic verification of C-D nozzle designs in aerospace propulsion, ensuring practical alignment performance with reliable results.

## METHODOLOGY

The approach for the quantitative and qualitative comparison between the Fluent and isentropic equations will utilize RPA-4 (rocket propulsion analysis) — a CEA (chemical element analysis) - based program for thrust chamber sizing and propulsion calculations. Three RPA example files will be used for the setup, with predetermined fuel constituent information, nominal thrust, and expansion and contraction area ratios, coupled with conical half angles. The latter parameters are essential for geometry. RPA-4 will generate a half-contour of the nozzle, creating a three-dimensional model for use in Fluent. RPA will additionally develop property variation across the geometry from the injector to the outlet plane. This data is then utilized inside ANSYS Fluent for CFD visualization. The CFD is set up to operate under steady-state conditions, with the energy equation solved along with the k-omega SST model. Additionally, the default Fluent material air is selected, with viscosity set to obey Sutherland's law for an ideal gas with variable density. Using RPA values, far-field pressure is set with default turbulent intensities of 5% and 10% and a viscosity ratio of 10. In this scenario, the outlet uses a backflow pressure specification based on the static pressure of each neighboring cell. At the same time, the walls are treated as no-slip with a standard roughness model. The numerical solution method is a coupled scheme with second-order upwind spatial discretization for all quantities except the gradient, using a least-squares cell-based discretization.

## RESULTS AND DISCUSSION

The following section describes the comparative analysis of three nozzle geometries. The first being the A-1 nozzle with a classical conical bell profile. The second RL10B2 has a dual bell nozzle with a breakpoint of approximately 1-3° at 60-70% of the cone length. Lastly, the SSME40k conical bell nozzle has a characteristic long thrust chamber. These geometries were generated using the Rocket Propulsion (RPA-4) program, which serves as a tool for combustion analyses and the conceptual design of nozzles with conical and parabolic profiles. The primary information required to begin conceptual construction of the thrust chamber is nominal thrust, nominal mass flow rate, and nozzle throat diameter. The nominal parameters (Table 1), along with fuel constituents and area ratios, define the contour used for three-dimensional modeling in SolidWorks (Figure 1). The nozzles can be used for CFD (computational fluid dynamics) modeling supported by Fluent. Values extracted from Fluent will be compared with user-defined code descriptions that follow classical one-dimensional isentropic flow equations [6].

**Table 2** Nozzle Parameters

| A-1 RPA Nozzle Parameters | |
|---|---|
| Property | Value |
| Engine Nominal Thrust (lbf) | 74700 |
| Ambient Pressure (atm) | 1 |
| Contraction Area Ratio | 1.6 |
| Expansion Area Ratio | 14 |
| Chamber Length (in.) | 45 |
| Contraction Angle b (degrees) | 20 |

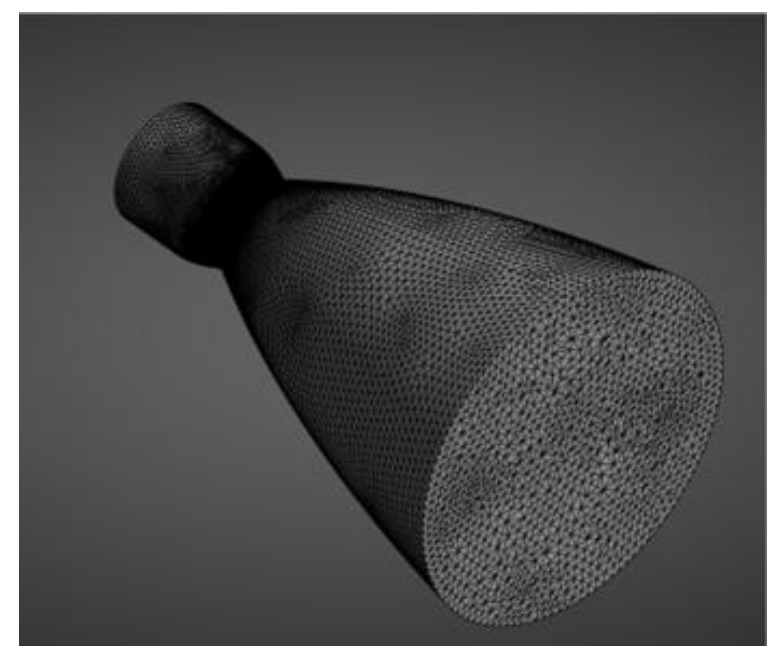

**Fig. 1** A-1 Nozzle Geometry

From one-dimensional flow equations, throat (characteristic) and exit (outlet) information can be defined. This includes pressure, temperature, density, velocities, and mass flow rate. These quantities will serve as the primary source of comparison against Fluent-procured quantities after meshing (Figure 2) and boundary condition setup. Evaluating these quantities will provide insight into the fluid behavior in each nozzle under specific engine operating conditions.

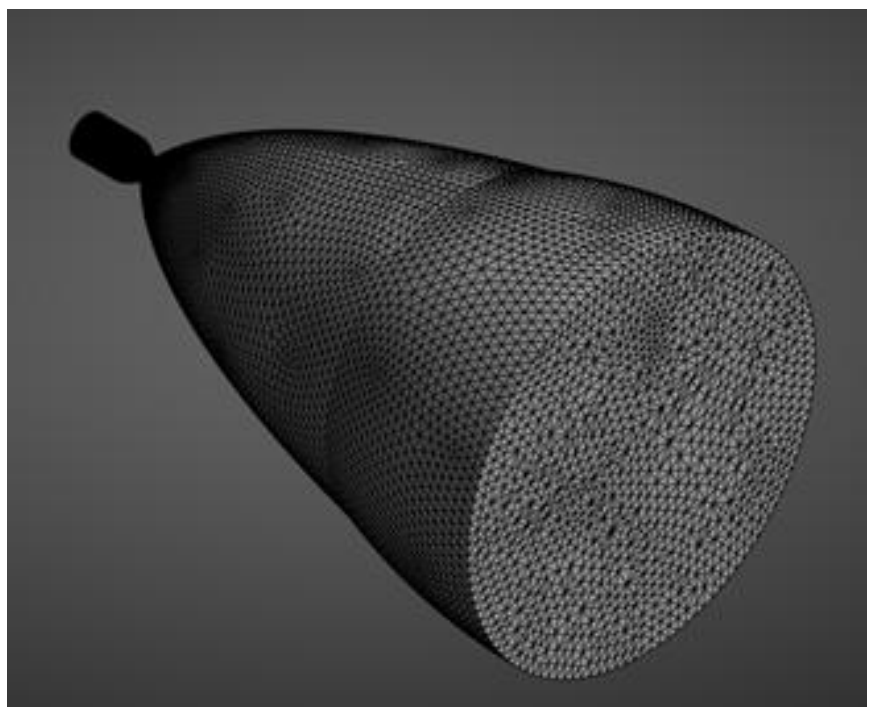

**Fig. 2** RL10B2 Mesh & Geometry

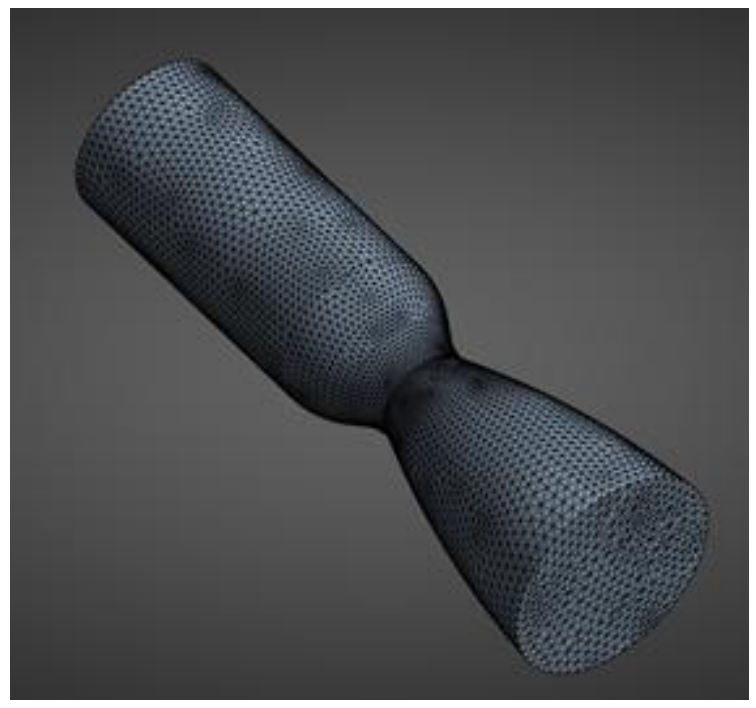

**Fig. 3** SSME 40k Mesh & Geometry

The primary metric for visualizing flow regime transition was the use of Fluent contours. From observation, some noise and blurriness are due to limited computational resources. However, Mach contours not only reveal transitions but also demonstrate that calculated RPA quantities can be used to establish boundary conditions for examining important fluid phenomena. The A-1 (Figure 4) classical conical bell nozzle exhibits a near-continuous supersonic expansion with minimal flow separation. Separation from walls can lead to turbulence consequences, including unaccounted ablation and heat transfer mechanisms. The same is true for the SSME 40K nozzle (Figure 5); although it shares a long-thrust chamber, the SSME 40K's larger expansion ratio shows a strong supersonic core. The observations between the first two nozzles align with nozzle configurations for altitude comparison [7] [8]. To some degree, however, the opposite is true of the RL10B2 nozzle (Figure 6), where some flow transitions occur. Upon closer inspection, the coarse mesh may have had a significant impact near the inflection region, where some flow amplification is present. In the diverging region, the coarse mesh may have led to under-resolved boundary layers and lip regions developing recirculation zones, resulting in lower-Mach areas than those that may have been physically present. Additionally, the SST k-ω model may have produced an artificial recirculation region and may not have been the most optimal turbulence method for capturing actual transient dynamics.

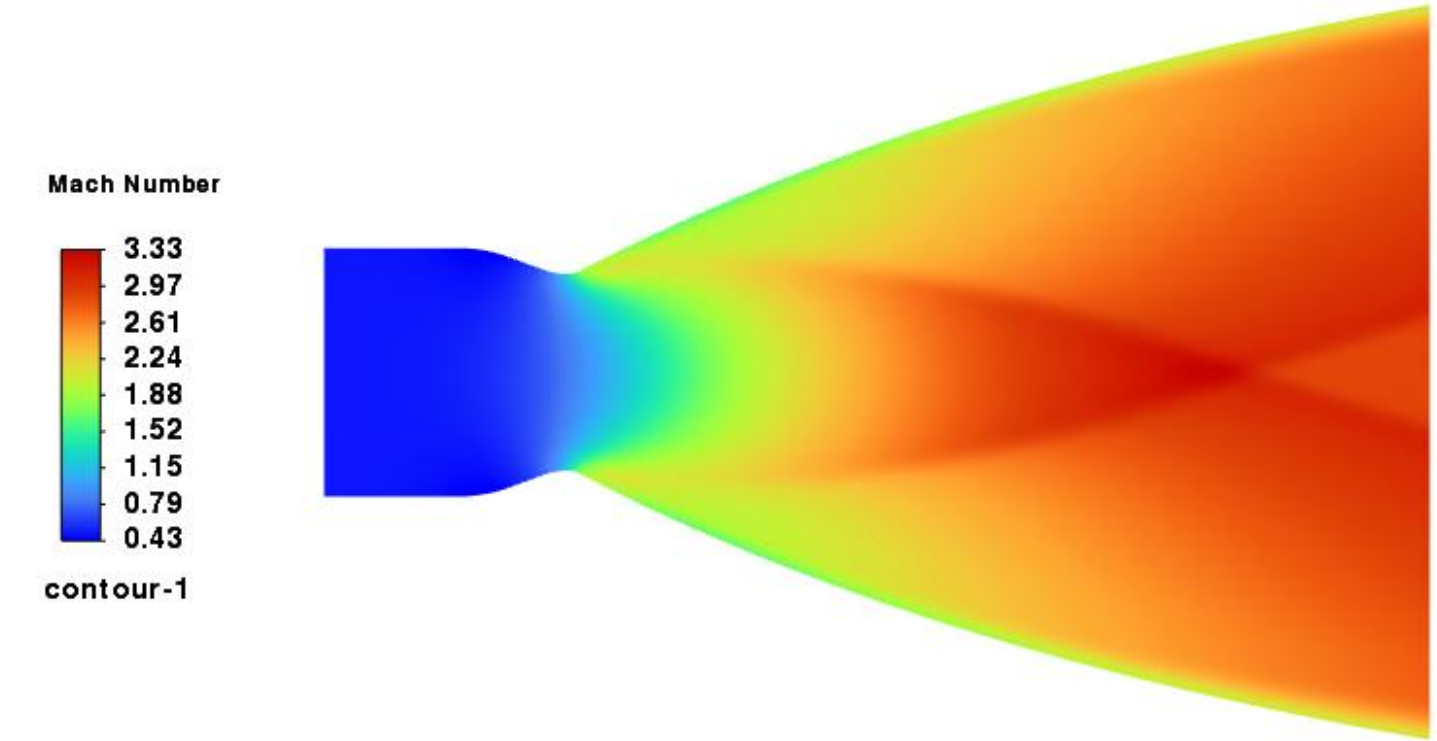

**Fig. 4** A-1 Mach Contour

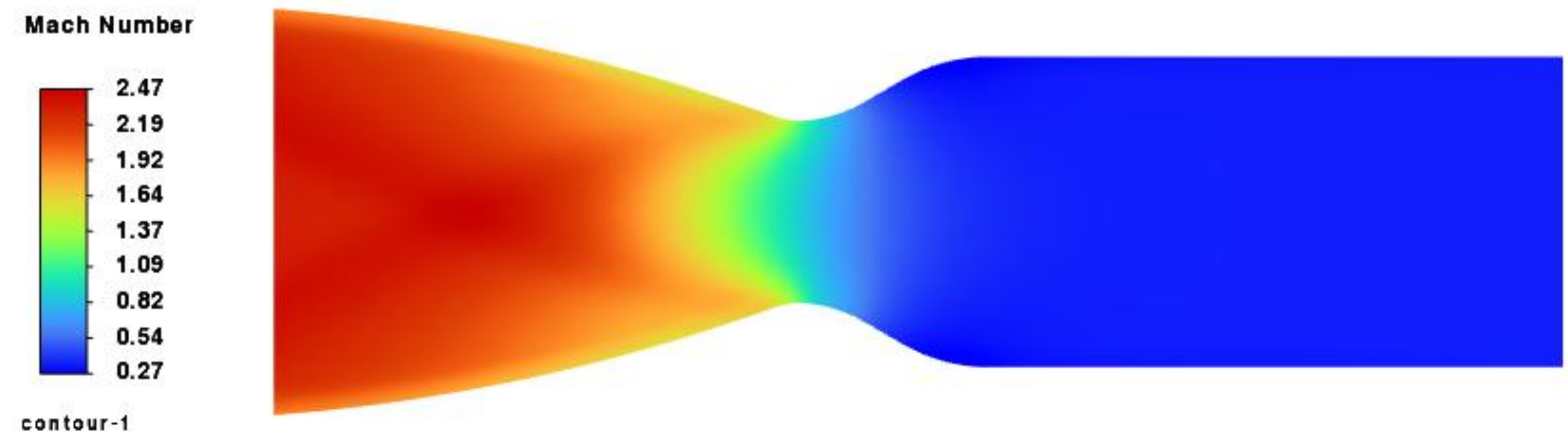

**Fig. 5** SSME 40K Mach Contour

For primary comparison against isentropic equations, an area-weighted average for the respective equations was extracted at a plane created normal to the flow at the inflection plane and outlet plane. The primary assumption in calculations is choking flow at the throat. This assumption is crucial for identifying the region where the flow regime transition occurs, and some adjustments may have inadequately approximated this location. Stagnation at the inlet, indicative of low velocity, is caused by combustion. In conjunction with the relationship between pressure and velocity under incompressible inviscid flow assumptions, the static pressure across all three nozzles exhibits the same fluid behavior. The contour in the inlet to the throat is defined as the region of highest pressure. This pressure characteristic is also used in isentropic calculations assuming Mach 1 at the throat, where the ratio values* are derived from one-dimensional compressible flow functions for an ideal gas with a ratio of specific heats equal to 1.4.

From an example calculation for the SSME 40k nozzle, theoretical calculations yield a throat pressure of 80.4 MPa to an exit pressure reduction of 0.53 MPa. Therefore, the pressure behavior between Fluent numerical approximations and isentropic calculations is consistent. Additionally, the fluid behavior due to pressure gradients near the wall region is apparent in all nozzles, indicative of boundary-layer interaction and potential separation. A similar effect, previously discussed [9] [10] in the context of dual-bell nozzle interaction, can also be observed in the RL10B2 nozzle (Figure 6).

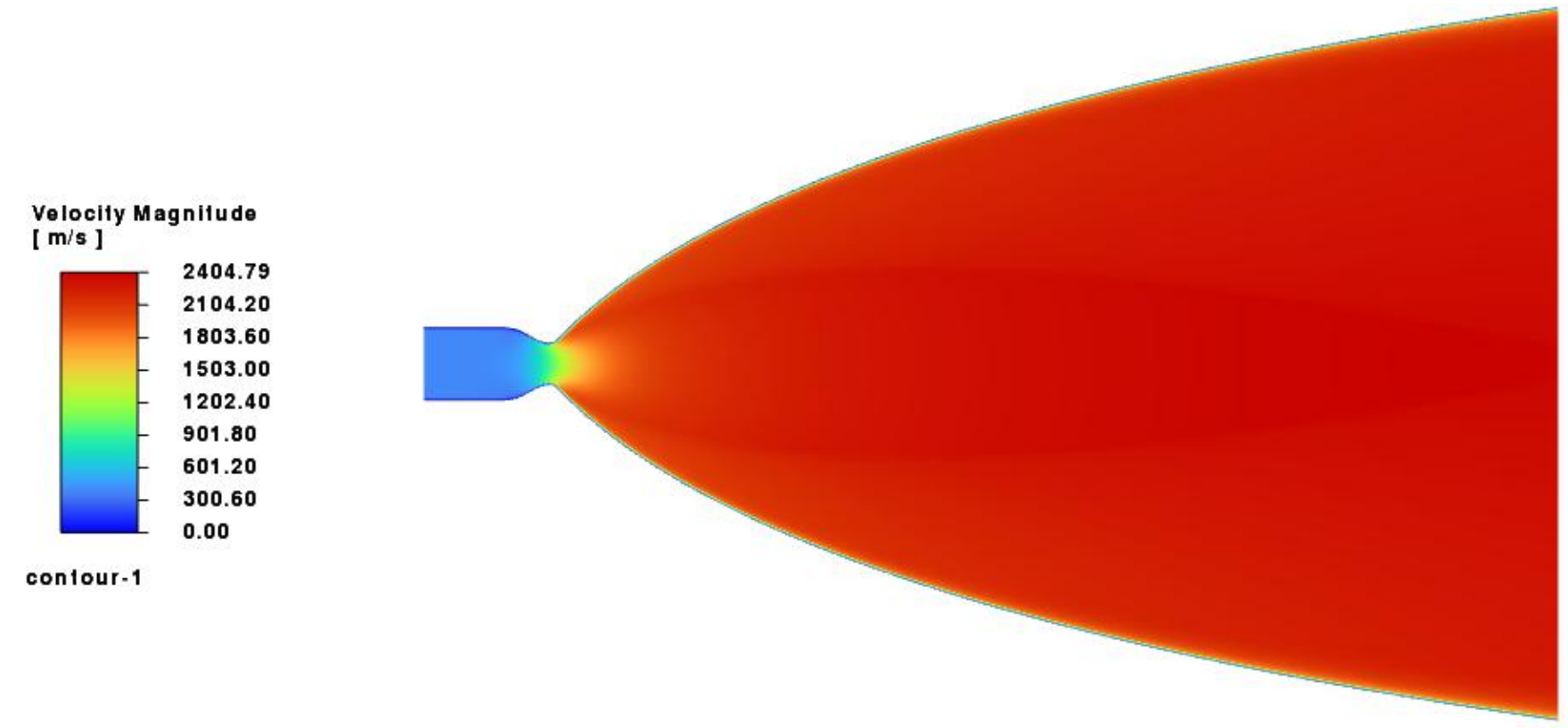

**Fig. 6** RL10B2 Boundary Layer Behavior

To assess flow acceleration and thermal loading, temperature and density distributions were used for each configuration. First, the A-1 nozzle demonstrated relatively uniform thermal gradients, suggesting efficient heat dissipation along the chamber walls. Conversely, the RL10B2 nozzle exhibited higher temperature gradients along the inner contour, particularly in the throat and near the shock inflection regions. Examination of the density contours (Figure 7) confirmed strong refraction at the nozzle exit. By examining theoretical calculations, the density refraction can also be discussed, with the SSME 40K nozzle density changing from 10.06 to 1.44 $kg/m^3$. This temperature-density coupling observed in Fluent is consistent with theoretical relationships derived from isentropic flow equations and with experimental results by Hadin and Oschwald [10].

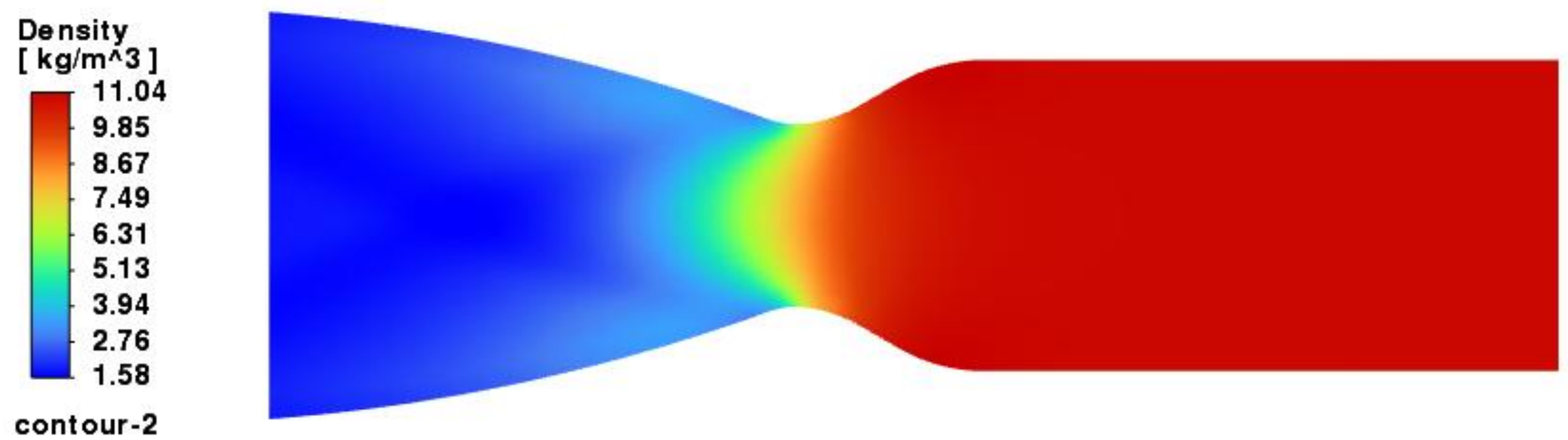

**Fig. 7** SSME 40K Density Contour

The three nozzle configurations show different levels of agreement between 1-D isentropic theory and SST RANS CFD (Figure 9).

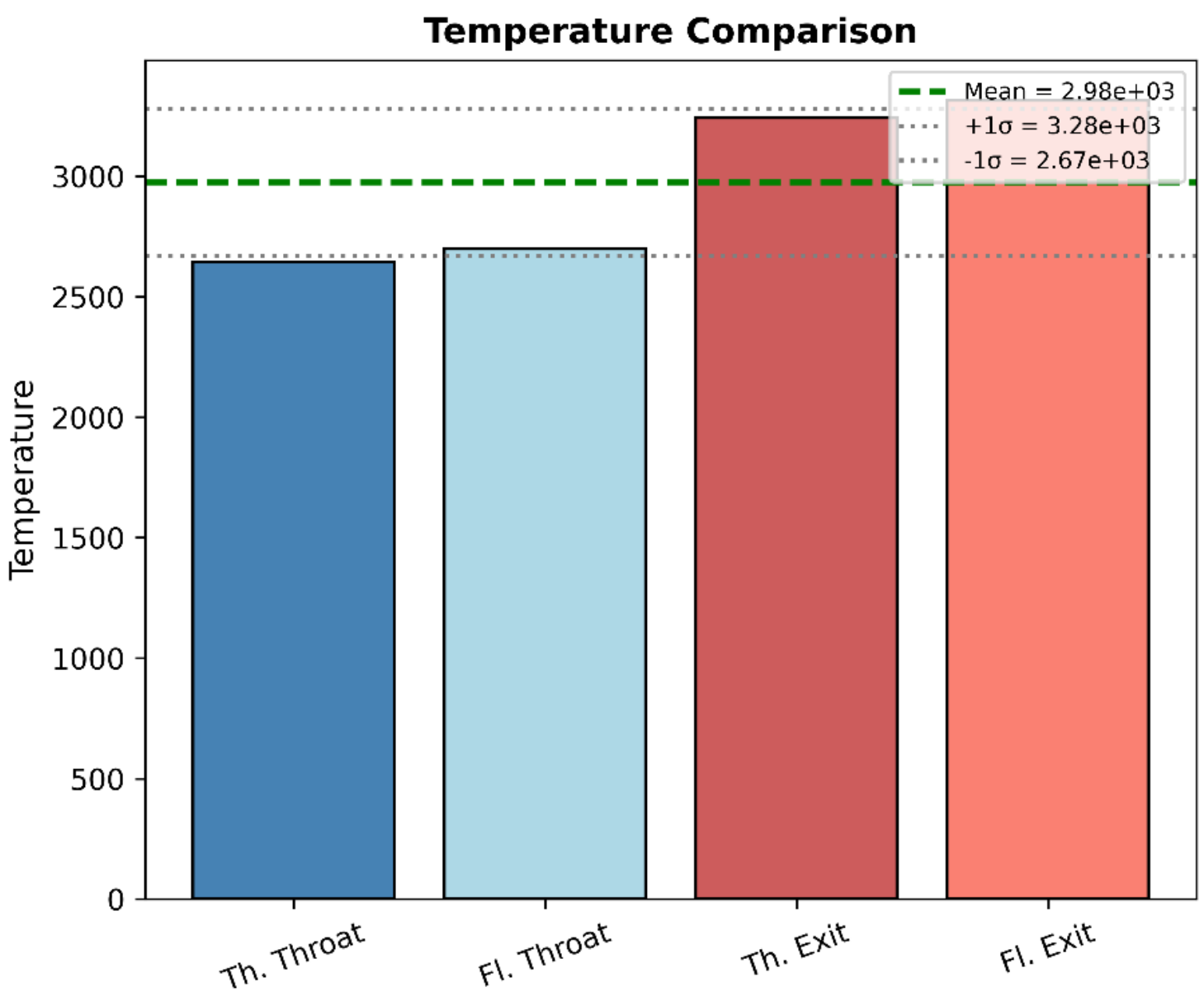

**Fig. 8** Temperature Comparison in the RL10B2 Nozzle

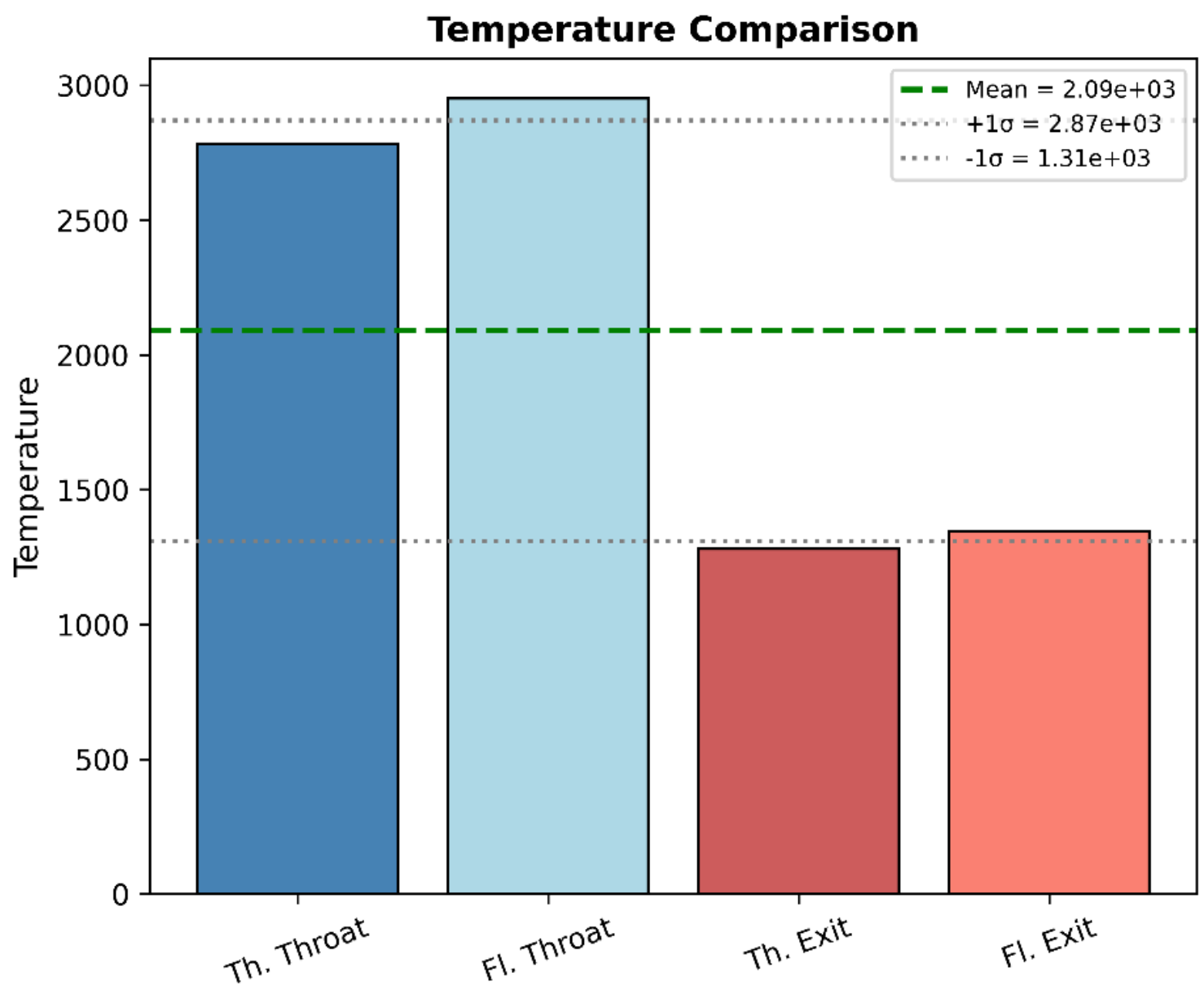

**Fig. 9** Temperature Comparison in the SSME-40k Nozzle

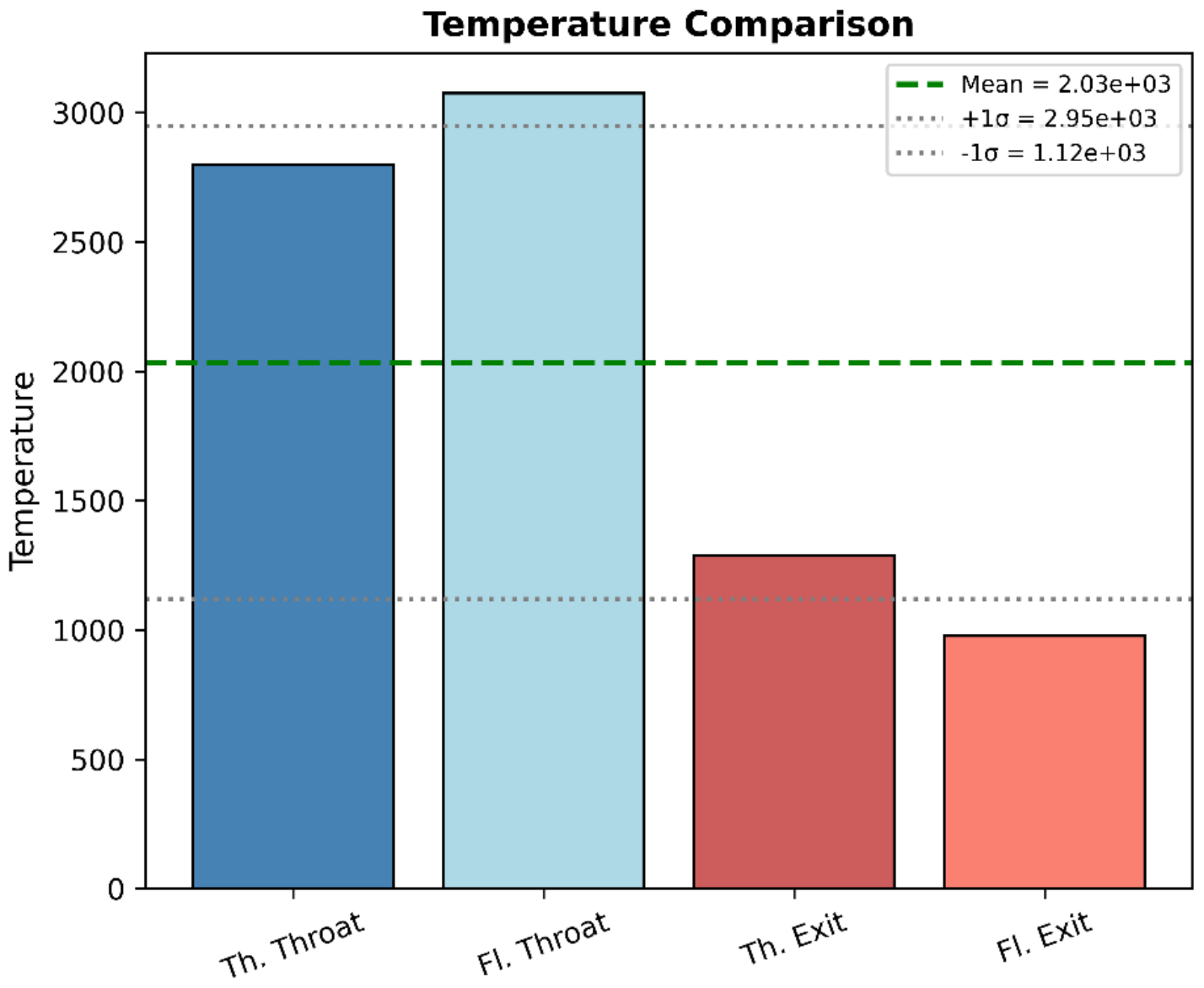

**Fig. 10** Temperature Comparison in the A-1 Nozzle

Consistent with viscous and three-dimensional stagnation recovery and boundary layer effects captured by the RANS SST k-ω model. Exit-plane deviations are larger and arise from non-uniform expansion and shock interactions; in particular, abnormal negative gauge pressures and low densities at the exit suggest either backflow or recirculation at the measurement plane, or misinterpretation of gauge versus absolute pressure. The RL10B2 dual-bell nozzle shows the most significant discrepancies. Where the steady RANS converged to a separated mode in which the inflection acted as a partial choke, producing near-zero exit pressure and low density, this behavior is expected for dual bell designs near transition. It indicates that transient mode switching, or refined meshing, is required to capture the nozzle's proper operation.

## CONCLUSION

The comparative analysis of the A-1, RL10B2, and SSME40K nozzle configurations demonstrated how geometric design fundamentally influences internal flow characteristics, pressure adaptation, and thermal behavior under high-speed propulsion conditions. The integration of isentropic one-dimensional theory with three-dimensional CFD analysis in Fluent provided a robust framework for assessing the fidelity of classical analytical methods against modern computational tools. Results for the A-1 and SSME40K nozzles showed strong agreement with theoretical predictions, with throat and exit parameters varying by 10–25% due to viscous and three-dimensional effects captured by the SST $k$–$\omega$ turbulence model. These deviations highlight Fluent's ability to resolve boundary-layer dynamics and stagnation recovery effects that are not captured by isentropic approximations.

In contrast, the RL10B2 dual-bell nozzle exhibited distinct flow separation and partial choking phenomena at the inflection region, reflecting its inherent sensitivity to pressure ratio and geometric breakpoint. The observed discrepancies between theoretical and numerical results in the RL10B2 case suggest that steady-state RANS modeling may inadequately represent transient flow switching behavior typical of dual-bell configurations. Refinement of the mesh and the incorporation of unsteady simulation techniques are recommended to better capture separation onset and shock-induced recirculation zones.

Thermal and density contours across all three nozzles revealed consistent temperature–density coupling, confirming the validity of isentropic relationships and experimental findings by Haidn and Oschwald [10]. The SSME40K nozzle demonstrated the most efficient expansion characteristics, maintaining stable supersonic cores with uniform heat distribution, while the A-1 profile provided a balanced compromise between simplicity and performance predictability.

Overall, this study underscores the importance of geometric tailoring in optimizing nozzle performance and pressure adaptation, particularly in high-altitude and dual-mode propulsion systems. The consistency between theoretical and computational analyses supports the continued use of RPA-generated geometries and Fluent-based CFD as reliable design and diagnostic tools for future nozzle development. Future work should extend to transient, multi-phase, and high-temperature reactive flow modeling to more accurately capture real propulsion environments and improve predictive capability for advanced rocket nozzle configurations.

### ACKNOWLEDGEMENT

This project utilized Prof. Arturo Rodriguez's Startup funds, which were granted by Texas A&M University-Kingsville, the Department of Mechanical and Industrial Engineering, and the College of Engineering. This project was also supported by the DOE NNSA/MSIPP Grande CARES Consortium (GRANT13584020).